# Topological-transition-driven Giant Enhancement of Second-harmonic Generation in Ferroelectric Bismuth Monolayer

Wen-Zheng Chen[1], Hongjun Xiang[1,*], Yusheng Hou[2,*]

[1] Key Laboratory of Computational Physical Sciences (Ministry of Education), Institute of Computational Physical Sciences, State Key Laboratory of Surface Physics and Department of Physics, Fudan University, Shanghai 200433, China

[2] Guangdong Provincial Key Laboratory of Magnetoelectric Physics and Devices, Center for Neutron Science and Technology, School of Physics, Sun Yat-Sen University, Guangzhou 510275, China

## Abstract

The interplay between band topology and light in condensed materials could unlock intriguing nonlinear optical phenomena, enabling modern photonic technologies such as quantum light sources and sub-wavelength topological lasers. Here, we unveil that a buckling-tuned topological transition in ferroelectric bismuth monolayer unleashes a giant second-harmonic generation. Using first-principles calculations, we surprisingly find that ferroelectric bismuth monolayer with a buckling parameter, $\Delta h$, has a large susceptibility $\chi^{(2)}$ on the order of $10^7$ $pm^2/V$, exceeding monolayer $MoS_2$ by about two orders of magnitude. When $\Delta h$ is engineered to the critical window where Dirac electrons emerge, a low-frequency resonance appears, boosting $\chi^{(2)}$ by an additional order of magnitude. We show that this enhancement is localized on the Dirac cones and dominated by intraband modification contributions. Based on an extended Dirac model, we establish that this enhancement physically originates from the ultralight effective masses $m^*$ of Dirac electrons through scaling with the Fermi velocity $v_F$ and band gap $E_g$. Our findings provide a viable route to achieving exceptional second-harmonic generation via engineering topological criticality, and could serve as an experimental signature of Dirac electrons in topological materials.

* Corresponding author: hxiang@fudan.edu.cn, houysh@mail.sysu.edu.cn

# 1. Introduction

Nonlinear optics (NLO) is a cornerstone of modern photonics, setting the physical basis for light frequency conversion, ultrafast light-matter control, on-chip modulation/detection, and nonlinear imaging/sensing [1-8]. In particular, integrated nonlinear photonics highlights materials and device engineering to raise conversion efficiency, lower power, and realize dense on-chip integration [2-5]. Among nonlinear processes, second-harmonic generation (SHG) plays a pivotal role due to its high signal-to-noise ratio, symmetry or domain probing, and direct device applications. Within an electric-dipole (ED) approximation, SHG is described by $P_i(2\omega) = \chi^{(2)}_{ijk}(2\omega)E_j(\omega)E_k(\omega)$ and vanishes in centrosymmetric crystals. Two-dimensional (2D) materials offer unique advantages for achieving large SHG, as their NLO susceptibilities, $\chi^{(2)}_{ijk}$, can rival or even exceed those of widely used bulk nonlinear materials by orders of magnitude [9-11]. In addition, their atomic thickness makes 2D materials highly amenable to interfacial and heterostructure engineering, enabling programmable SHG in TMDs, graphene, and h-BN systems [5, 6, 12-16].

Up to now, finding materials with large SHG remains a continuing challenge. It is pointed out by a famous empirical rule that $\chi^{(2)}$ links to the products of the linear susceptibility, $\chi^{(1)}$ [17, 18]. Exactly, ferroelectric (FE) materials typically have large $\chi^{(1)}$ and NLO responses have recently been reported in 2D ferroelectrics [11, 19-21]. So, intrinsically inversion-broken FE materials are promising for realizing giant SHG. On the other hand, owing to their unique band structures and topological features, topological materials provide platforms for exploring SHG and higher-order NLO responses. Nowadays, numerous efforts have been devoted to understanding the effect and physical mechanisms of band topology on NLO responses. For example, it is theoretically proposed that NLO responses can come from quantum geometric and Berry curvature dipole [22-24]. Giant susceptibility $\chi^{(2)}$ is pursued in Dirac/Weyl semimetals via induction mechanisms, exemplified by current-induced SHG [25-29]. Besides, higher-order effects beyond SHG have been studied to probe the interplay between topology and NLO [30, 31]. Even, experiments in TaAs confirm that linear band dispersion can yield record-high $\chi^{(2)}$ [32]. These studies hint a deep connection between Dirac dispersions and SHG responses [28, 29, 33, 34]. Because the inversion symmetry in ideal Dirac fermions restrict resonant enhancement to multipolar (i.e., non-ED) processes, previous studies almost focused on EQ-SHG/odd-order effects [30, 31, 35-38] or externally induced SHG [25-29]. Instead, the intrinsic ED-SHG is neglected [39]. As a result, an understanding of the intrinsic ED-SHG arising from band topology is missing.

In this context, Group-V single-element (i.e., As, Sb, Bi) monolayers (MLs) with the black-phosphorus (BP) structure are highly suitable starting systems. They exhibit tunable electronic structures, attractive for multifunctional optoelectronic, spintronic, and sensing applications [11, 40-43]. Notably, experiments verified that BP-Bi ML is an intrinsic 2D FE material and its polarization is strongly correlated to its buckling, $\Delta h$, which breaks inversion symmetry [44, 45]. Besides, BP-Bi ML has been grown on multiple substrates, and its $\Delta h$ is substrate-tunable [45-50]. Remarkably, it exhibits a topological phase transition tuned by both buckling engineering and FE control [46, 51]. Hence, via growth engineering and FE control, BP-Bi ML provides an ideal platform to explore how topological bands influences SHG in 2D FE systems.

In this work, we study the SHG of BP-Bi ML using first-principles calculations with spin-orbit coupling being included in all cases. We find a large ED-SHG arising from the intrinsic symmetry breaking and small band gap. Surprisingly, its $\chi^{(2)}$ exceeds that of ML $MoS_2$ by about two orders of magnitude. Remarkably, tuning $\Delta h$ near the topological phase transition produces a further order of magnitude enhancement in SHG. Via first-principles calculations and theoretical modeling, we trace this giant enhancement to the giant Fermi velocity, $v_F$, and small gap, $E_g$, characteristic of Dirac electron. We also provide a semiclassical picture in terms of the ultralight band-edge effective mass $m^*$ of Dirac carriers for understanding the enhancement. Therefore, BP-Bi ML unifies ferroelectricity-enabled ED activity and topological resonance amplification. Our work places BP-Bi ML among the strongest SHG materials and points to a band-topology-guided route to the giant $\chi^{(2)}$.

## 2. Buckling-controlled topological phase transition in ML BP-Bi

BP-Bi ML adopts a black-phosphorus-like lattice as shown in Fig. 1a [47]. Due to the ultra-heavy Bi and weak $6s$-$6p$ hybridizations, the bonding acquires partial $sp^2$ character rather than the homogeneous $sp^3$ configuration of black phosphorus [11, 45]. Such bonding produces a buckling $\Delta h = d_0$, which breaks inversion symmetry and changes the crystal symmetry from Pmna space group forbidding SHG to the polar $Pmn2_1$ one [46]. Our first-principles calculations further confirm the key structural and electronic features of BP-Bi ML in agreement with Ref. [45]. As shown in Fig. 1b, the inversion symmetry breaking allows two FE domain states with $\Delta h = \pm 0.45$Å, separated by a small double-well barrier. Buckling enlarges the band gap and lifts the sublattice $p_z$ degeneracy near Γ (Fig. S2a). For a given FE domain, the valence-band maximum derives mainly from $p_z$ orbitals on sublattice A and the conduction-band minimum from $p_z$ orbitals on sublattice B. Consequently, this yields a charge transfer from sublattice B to sublattice A and a spontaneous FE polarization which is nearly linear in $\Delta h$. This charge transfer is directly illustrated in the band structure of Fig. S1a, where BP-Bi ML exhibits an indirect band gap of 0.266 eV, consistent with scanning-probe measurements [45] and computational studies [52]. This agreement with experiment [45] confirms the low-energy band alignment underlying our SHG calculations. Projected density of states in Fig. S2b shows distinct peak splitting near $E_F$ in $p_z$ orbitals of sublattice A and B atoms, with sublattice A-derived states peaking below $E_F$ and sublattice B-derived states above $E_F$. The integrated density of states in Fig. S2b also reveals higher electron occupations in $p_z$ orbitals of sublattice A versus sublattice B. All these features establish $\Delta h$ as a FE order parameter of BP-Bi ML.

The band topology of BP-Bi ML is governed by its buckling, and adjusting this buckling can pass through trivial-topological phase transition boundaries. It has been demonstrated that the black-phosphorus-like "flat" phase is a topological insulator with a nontrivial $Z_2$ invariant and helical edge states. As the buckling $\Delta h$ increases, the band gap closes at a critical $\Delta h_c \approx 0.09$ Å with forming Dirac cones and then reopens at larger $\Delta h$ in a trivial insulating phase [46]. The topological character can be switched by the buckling $\Delta h$, which in practice is controlled by substrate coupling and charge transfer. For instance, a nearly flat BP-like phase is synthesized on HOPG [45, 46] while a strongly buckled and distorted BP-like phase is grown on Si(111) [47]. As shown in Fig. 1b, buckling $\Delta h$ is established as a practical control knob for the band topology in BP-Bi ML. Hereafter, we treat $\Delta h$ as the single structural parameter that simultaneously determines FE and topological orders in BP-Bi ML, and elucidate the dependence of the band dispersion and second-order susceptibility $\chi^{(2)}$ on $\Delta h$.

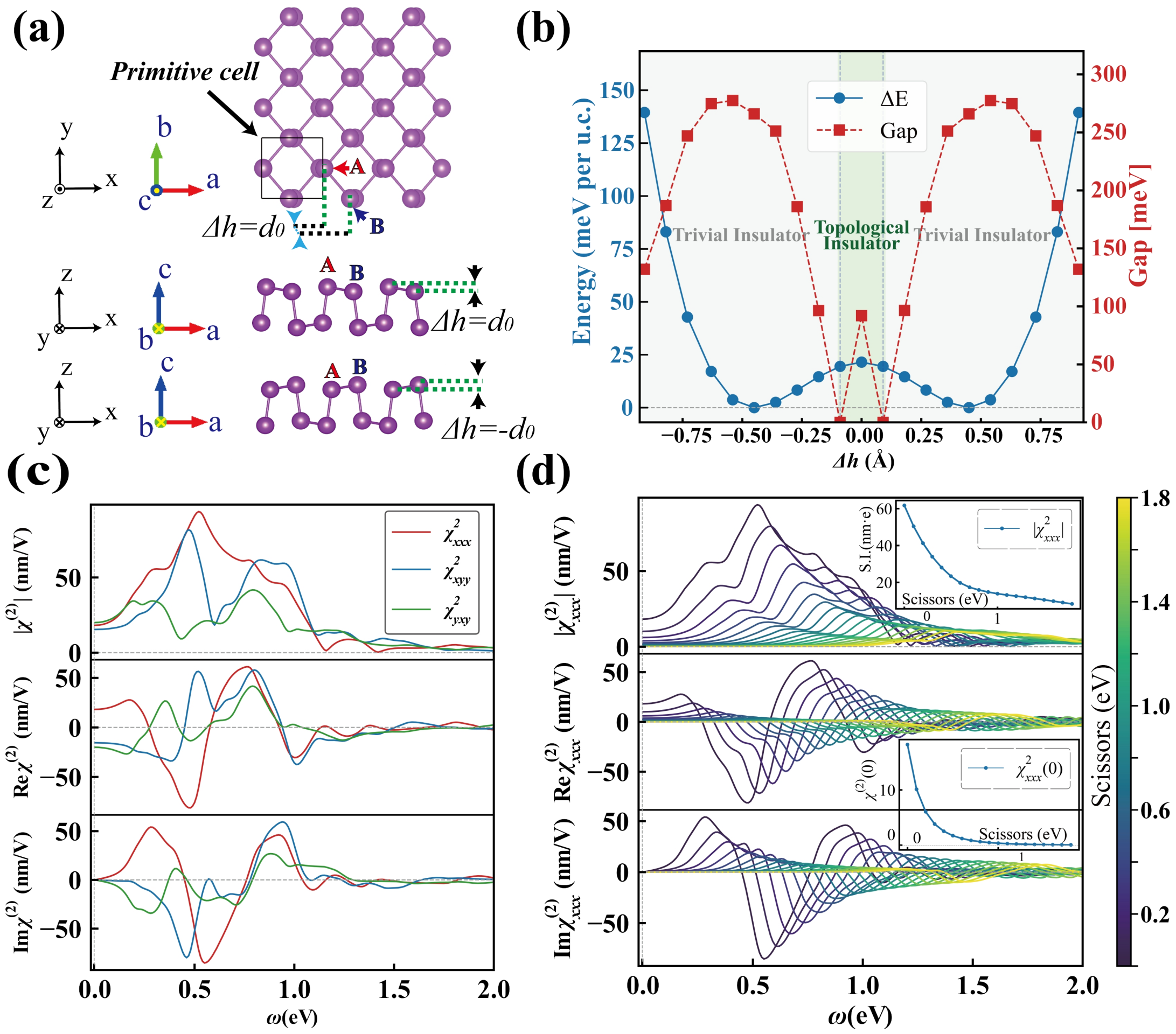

FIG. 1. Structural, electronic, and NLO properties of BP-Bi ML. (a) (top) top view and (middle) side view of BP-Bi ML with $\Delta h = d_0$; (Bottom) Side view of the $\Delta h = - d_0$ FE state. (b) Double-well potential, topological phase transitions, and band gap evolution with buckling $\Delta h$. (c) $\chi^{(2)}$ with their magnitude, imaginary, and real parts. (d) $\chi^{(2)}_{xxx}$ versus scissors-operator. Insets: (Top) spectral integrals versus scissors-operator. (Bottom): $\chi^{(2)}(0)$ versus scissors-operator.

## 3. Giant SHG in ML BP-Bi

Based on $Pmn2_1$ space group and $C_{2v}$ point group in BP-Bi ML [11], we obtain that its second-order susceptibility $\chi^{(2)}$ reduces to five independent nonzero components: $\chi^{(2)}_{xxx}$, $\chi^{(2)}_{xyy}$, $\chi^{(2)}_{xzz}$, $\chi^{(2)}_{yxy} = \chi^{(2)}_{yyx}$ and $\chi^{(2)}_{zxz} = \chi^{(2)}_{zzx}$. As BP-Bi ML is a 2D system, our analysis focuses on the in-plane tensor elements of $\chi^{(2)}$. Fig. 1c displays the SHG susceptibility $\chi^{(2)}$ curves of dominant components. We see that BP-Bi ML has a very large $\chi^{(2)}_{ijk}$, reaching the highest reported order of magnitude in 2D materials [5, 53]. Its main peaks' amplitudes reach the 10 nm/V order (see Section 4 of Supplement Materia for unit discussion) and concentrate below 1.5 eV, indicating strong infrared nonlinear responses. The dominant components, under the same computational framework, exceed those of ML $MoS_2$ by two orders, h-BN by three orders, and both polar metals (TaAs, TaP, NbAs) and conventional electro-optic materials (GaAs, ZnTe) by over one order of magnitude within the same frequency window [11, 21, 32]. Additionally, BP-Bi ML exhibits substantial static SHG responses $\chi^{(2)}(0,0,0)$ surpassing those of ML $MoS_2$, h-BN, and group-IV monochalcogenides [11, 21], which enables efficient, low-loss infrared frequency doubling for on-chip SHG.

The exceptionally large SHG in BP-Bi ML is fundamentally rooted in its narrow band gap. As thoroughly discussed in Section 2 of Supplement Material, all terms of $\chi^{(2)}(-2\omega;\omega,\omega)$ contain denominators $(\omega_{mn}-\omega)^{-1}$ and $(\omega_{mn}-2\omega)^{-1}$ which encode one- and two-photon resonances as shown in Fig. S1a. In these denominators, $\omega_{mn}=(E_m-E_n)/\hbar$ is the transition energy and $\omega$ represents the photon energy. These factors generate the peaks of the susceptibility and place their energies under direct control of the band gap through their formal divergences. A smaller gap reduces the relevant $\omega_{mn}$ and pulls the dominant resonances to low photon energies, which explains why the main peaks of BP-Bi lie below 1.5 eV and why the infrared response is strong. Consequently, the outstanding infrared response of BP-Bi ML arises from low-frequency one- and two-photon resonances. Besides $(\omega_{mn}-\omega)^{-1}$ and $(\omega_{mn}-2\omega)^{-1}$ terms discussed above, its giant SHG also follows directly from the additional $\omega_{mn}^{-1}$ factor. In length-gauge formulations, $\omega_{mn}^{-1}$ originates from the relation $\mathbf{r}_{mn}\propto\mathbf{v}_{mn}/\omega_{mn}$ between interband position and velocity matrix elements, and thereby quantifies the electronic compliance to the driving field [54]. As shown in Fig. S5, the small gap of BP-Bi ML leads to a large $\omega_{mn}^{-1}$ that increases the electronic compliance and contributes comparably to three different channels as detailed in Section 2 of Supplement Material. These channels correspond to the interband coherence, geometry-driven displacement and field-driven intraband motion. Their competition and cooperation raise low-energy spectral weight and enhance both the finite-frequency $\chi^{(2)}$ and static $\chi^{(2)}(0)$.

To make clear the gap contribution to SHG, we use the scissors-operator method which rigidly shifts the conduction-band manifold [55]. We adjust the gap from $E_g=0.266$ eV in BP-Bi ML to an artificial $E_g^{art}=2.0$ eV which is comparable to $MoS_2$

ML, and recompute $\chi^{(2)}$. As shown in Fig. 1d, $\chi^{(2)}_{\mathrm{xxx}}$ spectra (see Fig. S6 for other components) exhibit a clear monotonic trend. As $E_g$ increases, the dominant peaks move to higher photon energies and their amplitudes decrease systematically and by one order of magnitude at $E_g^{art} = 2.0\ eV$. A further analysis of the spectral integrals (see the inset in Fig. 1d) confirms the spectral amplitude reduction with increasing band gap. This behavior arises from the factors $(\omega_{\mathrm{mn}} - \omega)^{-1}$, $(\omega_{\mathrm{mn}} - 2\omega)^{-1}$ and $\omega_{\mathrm{mn}}^{-1}$ discussed above, which collectively reduce the low-energy spectral weight and overall spectral amplitude when $E_g$ increases. Consequently, both the finite-frequency $\chi^{(2)}$ and static $\chi^{(2)}(0)$ decrease. These results further confirm that the giant SHG of BP-Bi ML is essentially gap-driven. Hence, the small intrinsic $E_g$ places strong resonances in the near-infrared, enhancing the electronic compliance and maximizing their contribution to SHG.

Through Kramers-Kronig transformation, we can gain more insights into the physical origin of large static SHG responses of BP-Bi ML. The static $\chi^{(2)}(0)$ is an integral of the entire resonant spectrum and mathematically expressed as [11]:

$$\chi^{(2)}(0) = Re[\chi^{(2)}(0,0,0)] = \frac{2}{\pi}\mathcal{P}\int_0^\infty \frac{\mathrm{Im}\left[\chi^{(2)}(-2\omega;\omega,\omega)\right]}{\omega}d\omega \qquad (1).$$

Due to the $1/\omega$ kernel, the spectral weight of imaginary part $\mathrm{Im}\left[\chi^{(2)}(-2\omega;\omega,\omega)\right]$ at a low photon energy carries a larger contribution. In BP-Bi ML, the remarkably dense distribution of resonances below 1.5 eV (Fig. 1c), originating from its narrow band gap, creates a dominant spectral concentration region that governs this integral. This low-energy placement follows the same gap-controlled trend revealed by our scissors-operator analysis in Fig. 1d. As the gap increases, the peaks shift to higher energies and their amplitudes decrease, thereby lowering the static response. Crucially, the static susceptibility $\chi^{(2)}(0)$ (see inset in Fig. 1d) decays significantly faster than the spectral integrals. This behavior results from its dual sensitivity to both peak-position shifts and amplitude reduction, consistent with the kernel-weighting effect. In contrast to BP-Bi ML, wider-gap NLO materials such as $MoS_2$ and h-BN host their dominant resonances at higher photon energies. As a result, the $1/\omega$ kernel strongly suppresses their contribution to $\chi^{(2)}(0)$, thus accounting for much smaller static susceptibilities [21].

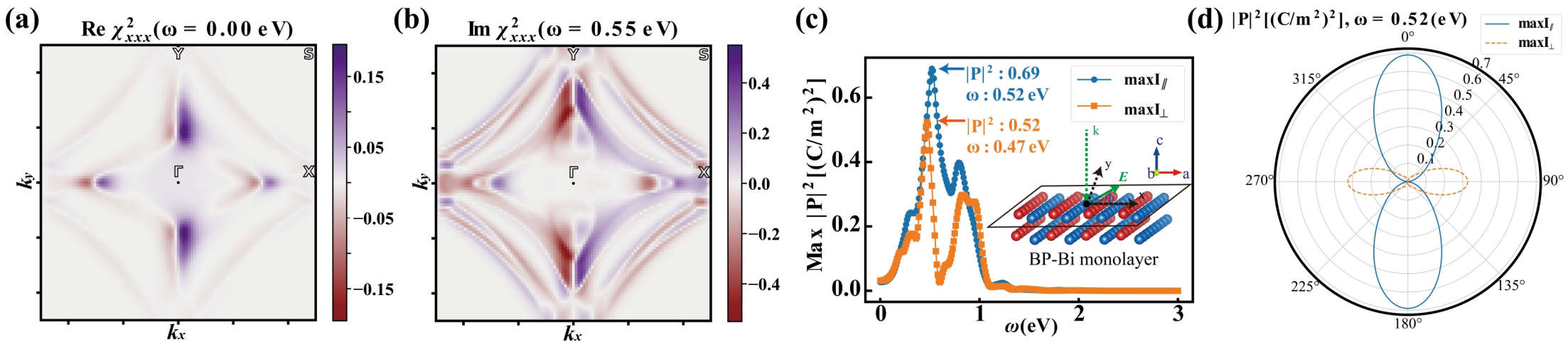

FIG. 2 $k$-resolved SHG responses and polarization dependence. (a) $k$-resolved static susceptibility $\chi^{(2)}_{xxx}(0)$. (b) $k$-resolved Im $\chi^{(2)}_{xxx}$ with an incident photon energy of 0.55 eV. (c) Maximum SHG intensity for co-polarized and cross-polarized configurations. (d) Polarization-resolved SHG with an incident photon energy of 0.52 eV.

We present Brillouin-zone-resolved contributions and polarization-dependent intensity of SHG in BP-Bi ML. Fig. 2a displays the $k$-resolved distribution of static susceptibility $\chi^{(2)}_{xxx}(0)$ (see Fig. S7 for other components). Fig. 2b show $k$-resolved contribution at the photon energies corresponding to the main peaks Im$\chi^{(2)}$ as shown in Fig. 1c (see Fig. S7 for real part and other photon energies). These k-resolved patterns are symmetric about $k_x$ axis and the SHG signal arises from a competition between the two sides of $k_y$ axis. This is consistent with the direction of the spontaneous FE polarization and $C_{2v}$ symmetry of BP-Bi ML. The static response $\chi^{(2)}(0)$ concentrates in regions with narrow gaps, in agreement with Kramers-Kronig analysis. As the photon energy increases, larger portions of the Brillouin zone start to contribute, but the dominant weight remains anchored in the narrow-gap sectors. Interestingly, these sectors lie along $\Gamma - X$ and $\Gamma - Y$, coinciding with sectors where $p_z$-orbital manifolds split.

Based on the relation $I(2\omega) \propto |P(2\omega)|^2$ and linearly polarized excitations, we study SHG intensity and show the results in Figs. 2c-2d. Fig. 2c depicts the maxima of co-polarized and cross-polarized SHG under excitations with the electric field parallel to BP-Bi ML (see inset in Fig. 2c). Fig. 2d shows the polarization-resolved SHG at the peak photon energy of co-polarized part (see Fig. S8a for cross-polarized part) as indicated in Fig. 2c. The co-polarized emission exhibits lobes along 0° and 180°, whereas the cross-polarized emission peaks at 90° and 270°, with two channels which are distributed orthogonally. Fig. S8b presents the polarization-resolved static SHG, which strictly follows $C_{2v}$ and displays a richer multi-lobe pattern. Overall, these results provide benchmarks for future experimental measurements and device designs in nonlinear optical applications.

## 4. Topological-transition-driven Giant Enhancement of SHG

By continuously tuning the buckling parameter, $\Delta h$, we investigate the critical behavior of SHG near topological phase transitions in BP-Bi ML. We find that once $\Delta h$ enters a narrow window around a critical value, the low-frequency SHG rises sharply. As shown in Fig. 3a, linearly reducing $\Delta h$ first produces a slow increase of the SHG governed by band gap reduction. Surprisingly, when $\Delta h$ reduces to a certain threshold window ($\Delta h \approx 0.27 \rightarrow 0.18$ Å, see Fig. S19), SHG exhibits an anomalous surge, with both the low-frequency $\chi^{(2)}(2\omega)$ and static $\chi^{(2)}(0)$ increasing by more than a factor of two. The $\chi^{(2)}_{yxy}$ trace in Fig. 3a (see Fig. S9 for other components) shows that this increase manifests as the emergence of a new resonance peak in the low-frequency region. Interestingly, this dramatic enhancement just coincides precisely with the emergence of Dirac cones in band structures (Fig. S1b). When $\Delta h$ is tuned near the topological phase boundary ( $\Delta h_C \approx 0.09$ Å ) in Fig. S1d and exemplified by $0.125$ Å (see Fig. S10 for other $\Delta h$), the Dirac cones fully form (Fig. S1c). The formation of Dirac cones triggers an extraordinary order-of-magnitude enhancement in SHG as shown in Fig. 3b. A channel-resolved decomposition in Fig. 3c (see Fig. S11 for real parts) shows that the low-frequency surge is dominated by the "intra" term, which is defined as the modification of the interband linear susceptibility by intraband motion. A $k$-resolved contribution analysis in Fig. 3d (see Fig. S12 for real parts and other components) further confirms that the low-frequency SHG collapses into the two Dirac cones, with negligible contributions from other regions. So, the Dirac-cone band structure is the primary driving force of the giant SHG enhancement.

To further elucidate the physical origin of the giant SHG enhancement, we decouple the Dirac-electron contribution, by explicitly separating it from $\chi^{(2)}(2\omega)$ based on the following four-band Dirac model [28, 32, 39, 56-58],

$$H(k)_{base} = \hbar\left(v_x k_x \sigma_x \otimes I_s + v_y k_y \sigma_y \otimes I_s\right) \quad (2).$$

In Eq. (2), $k_x$ and $k_y$ denote the crystal momentum offset from the Dirac point, $v_x$ and $v_y$ are anisotropic Fermi velocities, $\sigma_i$ ($i = x, y$) are Pauli matrices acting on the sublattice pseudospin degree of freedom, and $I_S$ is an identity matrix that preserves spin degeneracy. Owing to symmetry constraints, ED-SHG is strictly forbidden by Furry's theorem for massless Dirac fermions described by Eq. (2) [59]. However, the Dirac cones are anisotropic rather than perfectly symmetric in BP-Bi ML. Hence, we go beyond the Dirac approximation in order to more comprehensively characterize Dirac fermion behavior. Building on the base Dirac Hamiltonian, we augment it with critical refinements incorporating anisotropic mass corrections [60, 61], spin-orbit coupling [62-65], tilt operators [39, 66], and quadratic momentum terms [56, 67-70]. Taking these together, the extended Hamiltonian, $H_{Dirac}^{ext}$, describing Dirac electrons in BP-Bi ML is in the following form:

$$H_{Dirac}^{ext} = H_{base} + H_{mass} + H_{SOC} + H_{tilt} + H_{quadratic} \qquad (3).$$

For Eq. (3), the detailed expressions and matrix forms of each term are provided in Section 5 of Supplement Material.

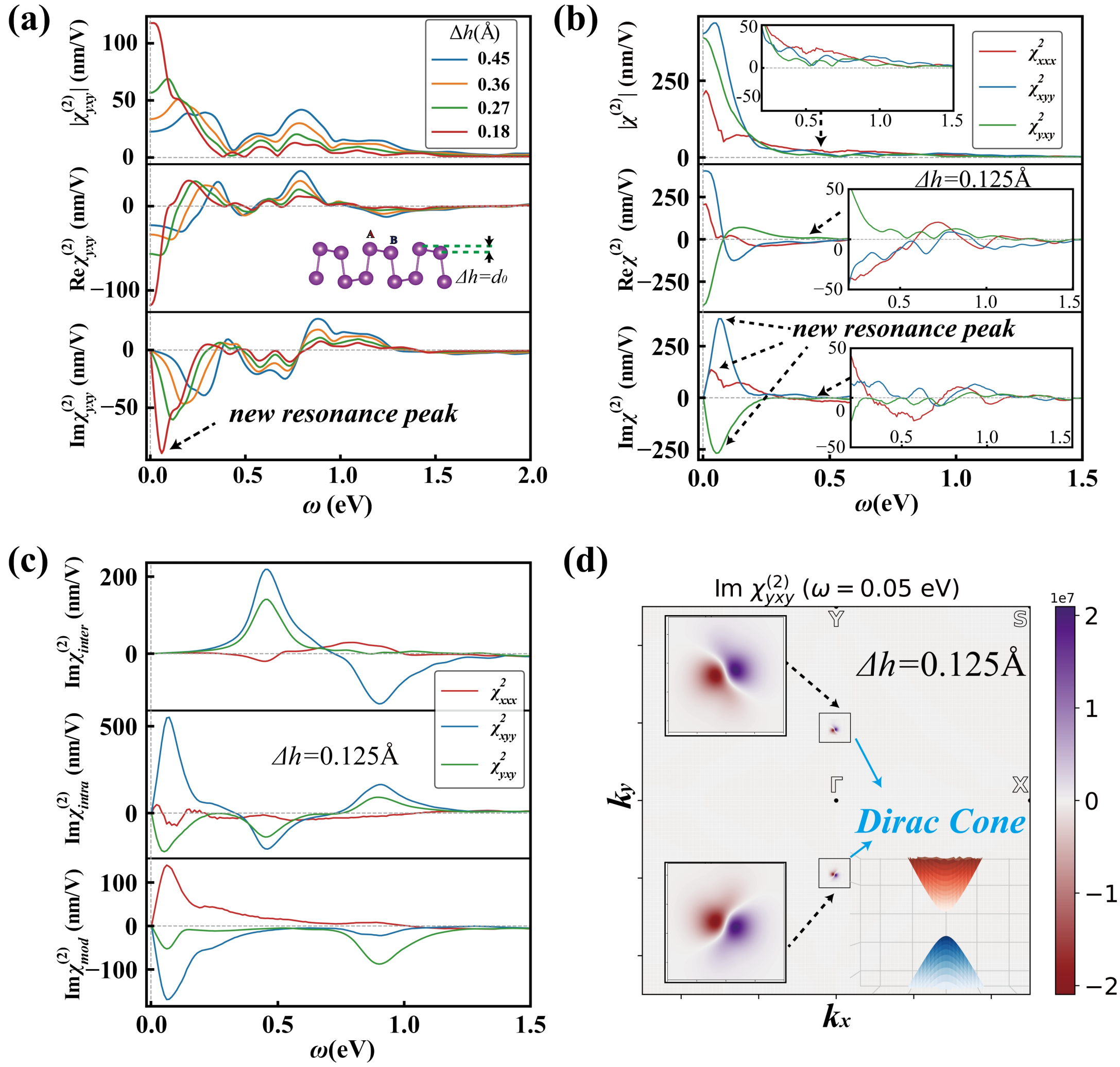

FIG. 3. SHG evolution with the topological phase transition in BP-Bi ML. (a) $\chi^{(2)}_{yxy}$

for several representative $\Delta h$. (b) $\chi^{(2)}$ at $\Delta h = 0.125$ Å. (c) Channel-resolved decomposition $\chi^{(2)}$ at $\Delta h = 0.125$ Å. (d) $k$-resolved contribution of the low-frequency SHG weight at the new resonance peak ω = 0.05 eV.

Based on $H_{Dirac}^{ext}$, we fit the Dirac cones of BP-Bi ML at $\Delta h = 0.125$ Å (Figs. 3d, S1c) and obtain $v_x \approx 3.42 \times 10^5\ m/s$ and $v_y \approx 2.42 \times 10^5\ m/s$. This fitted Hamiltonian reproduces the slopes, anisotropy, and the local dispersion near the cones, thus enabling a separate evaluation of the Dirac-cone induced SHG. Fig. 4a indicates that $\chi_{xyy}^{(2)}$ and $\chi_{yxy}^{(2)}$ components are concentrated entirely in the low-frequency region. They align with the new resonance as shown by the arrows in Figs. 3a-3b, nearly accounting for the whole enhancement of the low-energy SHG. Meanwhile, the $k$-resolved contribution of SHG obtained by the extended Dirac Hamiltonian in Fig. S15 shows excellent agreement with first-principles calculated results as shown in Fig. 3d and Fig. S12. This establishes that the surge of SHG near the topological transition is governed by the Dirac electrons. For $\chi_{xxx}^{(2)}$, our extended Hamiltonian yields a vanishing result because the fitted Dirac cone remains symmetric in the $k_x$ direction as verified in Fig. S14a. In this case, the relevant integrands are odd under $k_x \rightarrow -k_x$, which enforces cancellation over the Brillouin zone and suppresses the Dirac-cone contribution to $\chi_{xxx}^{(2)}$. Accordingly, the emergence of $\chi_{xxx}^{(2)}$ requires coupling to non-Dirac bands to lift the $k_x$-symmetry constraint. This dependence on non-Dirac bands explains why $\chi_{xxx}^{(2)}$ is markedly smaller than $\chi_{xyy}^{(2)}$ and $\chi_{yxy}^{(2)}$ in the regimes where SHG is enhanced by Dirac cones (Fig. 3b). This mechanism reverses the usual tensor hierarchy: $\chi_{xxx}^{(2)}$ is the dominant component when $\Delta h$ is in normal regime (Fig. 1c), but becomes the smallest as $\Delta h$ is in the Dirac-cone-enhanced SHG regime (Fig. 3b). Hence, our extended Hamiltonian reproduces the low-frequency peaks and its tensor selectivity. Moreover, it explains the order-of-magnitude enhancement in $\chi_{xyy}^{(2)}$ and $\chi_{yxy}^{(2)}$ while clarifying why $\chi_{\mathrm{xxx}}^{(2)}$ is comparatively weak in BP-Bi ML near the topological transition.

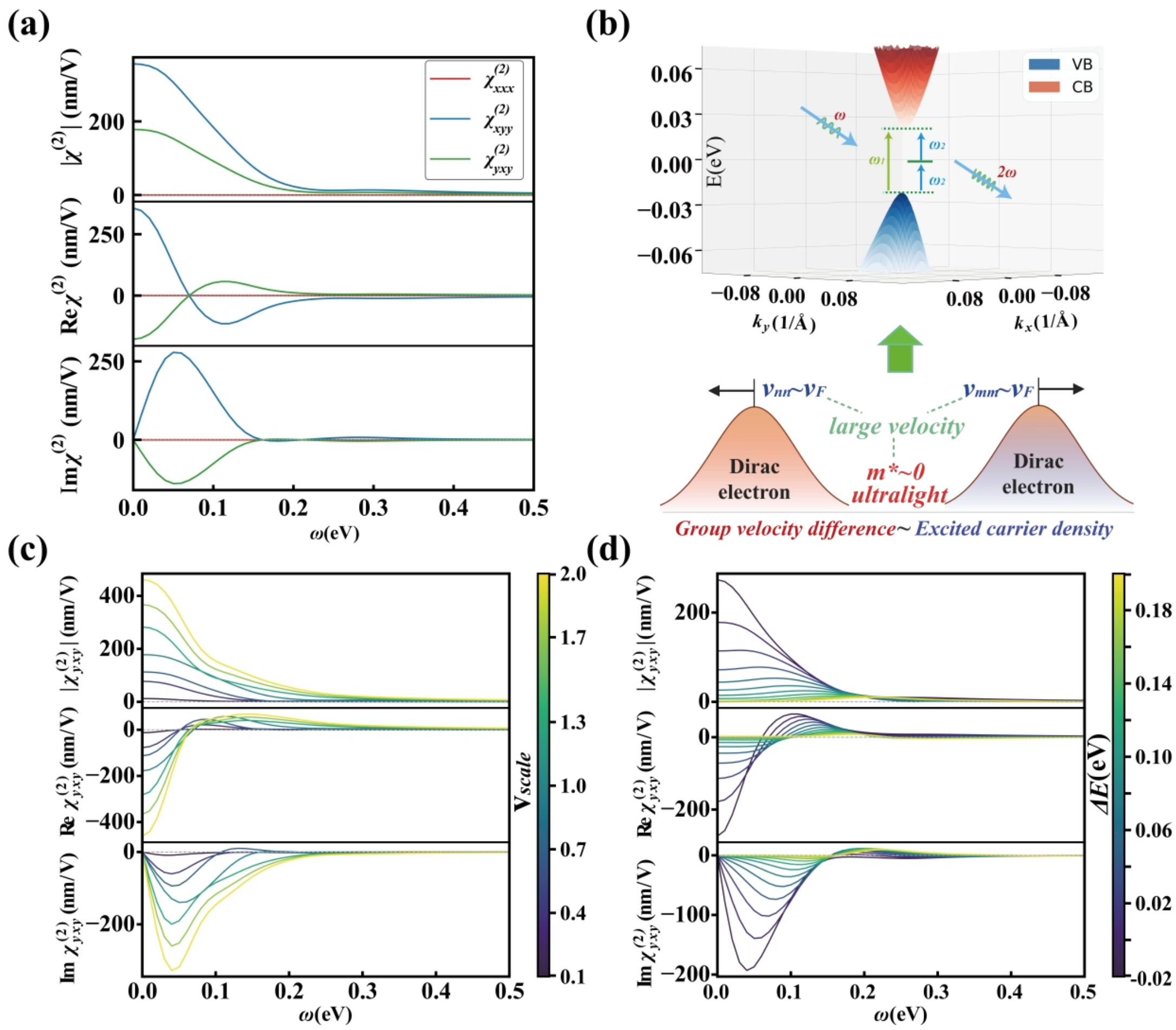

FIG. 4. Illustration of nonlinear scaling laws and the mechanism with the extended Dirac Hamiltonian. (a) Dirac-cone induced SHG spectra computed by extended Dirac Hamiltonian. (b) 3D Dirac cone viewed along the (110) direction and illustration of SHG enhancement mechanism. (c) Velocity scaling: SHG spectra computed for Fermi velocities scaled by a common factor $v_{scale}$, with $v_x' = v_{scale} \times v_x$ and $v_y' = v_{scale} \times v_y$. (d) Gap scaling: SHG spectra for band-gap offsets $\Delta E$ relative to the baseline structure.

To identify which features of Dirac electrons determine the magnitude of SHG, we extract SHG scaling relations with respect to the Fermi velocities and the mass gap, based on $H_{Dirac}^{ext}$. Figs. 4c-4d show that the low-frequency peak grows strongly with the Fermi velocities and is suppressed as the mass gap increases. This behavior follows the length-gauge formalism of $\chi^{(2)}$. In this formalism, the "intra" channel (Fig. S13, Eq. (S4)) contains velocity-difference tensors and $k$-covariant derivatives of the interband dipole, both of which scale with the group velocities. In addition, the small gap enhances the $\omega_{mn}^{-1}$ weighting, thus increasing the electronic compliance. Thus, large Fermi velocities and a small gap cooperatively maximize the Dirac-cone induced SHG. Actually, Dirac cones inherently have large local SHG amplitudes, but the total SHG is cancelled out in inversion-symmetric cones. In BP-Bi ML, Dirac cones possess an intrinsic tilt (Fig. S14b, S14c) unbalancing contributions and creating an asymmetric competition, which thereby reveals the otherwise hidden Dirac-cone induced SHG. Consistent with Fig. 3d, Fig. S12, and Fig. S15, the $k$-resolved weight concentrates on two Dirac cones with an asymmetric competition. This competition allows the intrinsically large Dirac-cone SHG, driven by large Fermi velocities and a small gap, to appear in the total SHG signal.

A complementary semiclassical perspective shown in Fig. 4b is also instructive to understand the giant enhancement of SHG in BP-Bi ML. At the Dirac point, which corresponds to both the conduction band minimum and valence band maximum, electrons acquire an exceptionally small effective mass $\mathrm{m}^* = \hbar^2\left(\frac{\partial^2 \mathrm{E}}{\partial \mathrm{k}^2}\right)^{-1}$. Such a small effective mass arises from the rapid energy variation across the cone apex. This can be made more explicit with a minimal gapped anisotropic Dirac model, $\mathrm{H}_{\mathrm{Dirac}}^{\mathrm{gap}} = \mathrm{H}_{\mathrm{base}} + \mathrm{H}_{\mathrm{mass}} = \hbar v_x k_x \sigma_{\mathrm{x}} + \hbar v_y k_y \sigma_{\mathrm{y}} + \Delta\sigma_{\mathrm{z}}$, whose eigenvalue bands are $\mathrm{E}_{\pm}(\mathrm{k}) = \pm\sqrt{\Delta^2 + (\hbar v_x k_x)^2 + (\hbar v_y k_y)^2}$ with $E_g = 2|\Delta|$. An expansion near the Dirac point gives a parabolic form for the conduction band $\mathrm{E}_{+}(\mathrm{k}) \approx \frac{\mathrm{E_g}}{2} + \frac{\hbar^2 k_x^2}{2\mathrm{m}_{\mathrm{x}}^*} + \frac{\hbar^2 k_y^2}{2\mathrm{m}_{\mathrm{y}}^*}$, with the anisotropic effective masses $\mathrm{m}_{\mathrm{x}}^* = \frac{\mathrm{E_g}}{2v_x^2}$ and $\mathrm{m}_{\mathrm{y}}^* = \frac{\mathrm{E_g}}{2v_y^2}$. Consistent with our earlier analysis, this effective mass expression reveals that the large Fermi velocities $v_F$ and small band gap $E_g$ are fundamentally embedded within the semiclassical framework through $\mathrm{m}^*$. At the Dirac point, this combination produces ultralight electrons (Fig. 4b). In semiconductor physics, material properties are typically governed by band-edge states, which here lie on the Dirac cones. As a result, the emergence of Dirac cones endows these electrons with extremely small effective masses, thereby strongly enhancing their electric-field susceptibility and amplifying SHG. Such enhancement from Dirac electrons is clearly resolved in the SHG intensity as shown in Fig. S16. In particular, the most prominent characteristic is an order-of-magnitude difference between co- and cross-polarized SHG in low-frequency regimes. This distinct signature establishes measurable criteria for experimental verification.

## 5. Conclusion

In summary, combining the first-principles calculations with an extended Dirac model, we establish BP-Bi ML as a ferroelectric topological platform with giant ED-SHG. The ferroelectric state together with a narrow band gap yields a large $\chi^{(2)}(2\omega)$ that exceeds $MoS_2$ ML by two orders of magnitude. Tuning the buckling $\Delta h$ near the topological transition produces a new low-frequency resonance and an additional order-of-magnitude enhancement in SHG. We show that the enhancement is localized on Dirac cones and dominated by intraband-modification contributions. Based on an extended Dirac model, we unveil that the SHG strongly increases with Fermi velocity and decreases with band gap. In a semiclassical picture, this scaling law follows from ultralight effective masses $m^* \propto E_g \big/ v_F^2$, linking the large Fermi velocity $v_F$ and the small gap $E_g$ of Dirac cones. Together with its mature synthesis techniques, our results suggest that BP-Bi ML is promising for low-power, on-chip near-infrared frequency conversion via substrate/strain induced buckling engineering and ferroelectric control. Importantly, our findings establish introducing topological quantum transitions in ferroelectrics as an appealing paradigm for achieving giant NLO responses.

**ACKNOWLEDGMENTS**

We acknowledge financial support from the National Key R&D Program of China (No. 2024YFA1408303, 2022YFA1402901), NSFC (grants No. 12474247, 12188101), Shanghai Science and Technology Program (No. 23JC1400900), the Guangdong Major Project of the Basic and Applied Basic Research (Future functional materials under extreme conditions--2021B0301030005), Shanghai Pilot Program for Basic Research-Fudan University 21TQ1400100 (23TQ017). Y. Hou acknowledges the support from Guangdong Provincial Key Laboratory of Magnetoelectric Physics and Devices (Grant No. 2022B1212010008) and Research Center for Magnetoelectric Physics of Guangdong Province (Grant No. 2024B0303390001).